\documentclass[aps,pre,showpacs,preprintnumbers,amssymb,amsfonts,floatfix,a4paper,twoside,twocolumn]{revtex4}
\usepackage[utf8]{inputenc}
\usepackage[T1]{fontenc}
\usepackage{lmodern}
\usepackage{amsmath,amssymb}
\usepackage{xcolor}
\usepackage{graphicx}
\usepackage[colorlinks,linkcolor=blue,urlcolor=blue,citecolor=blue]{hyperref}
\usepackage{mathptmx} 
\usepackage{xspace}
\usepackage{lipsum}
\DeclareMathOperator\erf{\mathrm{erf}}
\newcommand{\eref}[1]{Eq.~(\ref{#1})}%
\newcommand{\erefs}[1]{(\ref{#1})}%
\newcommand{\erefss}[1]{Eqs.~(\ref{#1})}%
\newcommand{\fref}[1]{Fig.~\ref{#1}} %
\newcommand{\ffref}[1]{Figure~\ref{#1}} %
\newcommand{\frefs}[1]{\ref{#1}} %
\newcommand{\frefss}[1]{Figs.~\ref{#1}} %
\newcommand{\sref}[1]{Sec.~\ref{#1}}%
\newcommand{\ssref}[1]{Section~\ref{#1}}%
\newcommand{\srefss}[1]{Secs.~\ref{#1}}%
\newcommand{\srefs}[1]{~\ref{#1}}%
\newcommand{\aref}[1]{Appendix}%

\begin{document}

\title{Stochastic efficiency of an isothermal work-to-work converter engine}

\author{Deepak Gupta}
\author{Sanjib Sabhapandit}
\affiliation{Raman Research Institute, Bangalore 560080, India}
\date{\today}
\begin{abstract}
We investigate the efficiency of an isothermal Brownian work-to-work
converter engine, composed of a Brownian particle coupled to a heat
bath at a constant temperature. The system is maintained out of
equilibrium by using two external time-dependent stochastic Gaussian
forces, where one is called load force and the other is called drive
force.  Work done by these two forces are stochastic quantities. The
efficiency of this small engine is defined as the ratio of stochastic
work done against load force to stochastic work done by the drive
force. The probability density function as well as large deviation
function of the stochastic efficiency are studied analytically and
verified by numerical simulations.
\end{abstract}
\maketitle
\section{Introduction}
\label{intro}
Heat engine \cite{Callen,Zemansky} is a machine that operates between
two temperatures in a cyclic process. It converts a part of the heat
$Q_H$ taken from the hot reservoir at a temperature $T_H$ to useful
work $W,$ and the remaining part of the heat $Q_C$ is dumped into the
cold reservoir at a temperature $T_C<T_H$. At the end of the cyclic
process, the engine returns to its initial state. The efficiency of an
engine is given by the ratio of work done by it to the heat consumed
from the hot reservoir: $\eta=W/Q_H$. When such engines
work in the quasi-static limit as well as in a reversible fashion, its
efficiency is given by the Carnot efficiency $\eta_C=1-T_C/T_H$. The
efficiency of any engine is bounded above by the Carnot efficiency: $\eta\leq\eta_C$. This bound is universal, and does not
depend upon the nature of the composition of the engine. In the
quasi-static regime, the power delivered by the engine is identically
zero: $W/t\ \to0$ in the limit $t\to \infty$. Therefore,
the Carnot engine is not useful for doing work in a reasonable time in
practice.

Modern technology helps in engineering machines on a microscopic
scale. These small nanosized devices can be seen in many areas of
biological science \cite{Bruce,Sugawa,Bio-1,Bio-2,Bio-3,Bio-4}. The
fluctuations present in the surrounding environment can disturb the
deterministic nature of such small-scale devices. Nevertheless, the
state of the system can be described in the probabilistic manner,
whose evolution is governed by the master equation or Fokker-Planck
equation. Interestingly, nowadays various properties of these small
systems can be understood by realizing them in controlled
experiments~\cite{Blickle,Martinez,Sudeesh,Saira,Toyabe,Ciliberto,Collin,Joubaud,Gomez}.

When such small-scale machines are driven by external forces, like
temperature or concentration gradient, shear flow, time-dependent
external field, etc., observables such as work done, heat flow, power
injection, entropy production, etc., become stochastic quantities
\cite{SS-1,SS-2,Apal-1,Apal-2,Verley-11,VanZon-1,VanZon-2,VanZon-3,Wang,Visco,Kundu,
  Farago2002, Deepak}. The probability distributions of these
quantities have richer information than their ensemble average values.

Over the past two decades, a lot of research has been devoted to
refining the thermodynamic principle in the mesoscopic scale. While the
first law of thermodynamics is also valid at the trajectory level, the
second law of thermodynamics is replaced by the symmetry property of
the probability distribution of total entropy production
\cite{Sekimoto,Seifert-1,Seifert-2}. This symmetry property is
referred to as the \emph{fluctuation theorem} (FT) \cite{Evans-Cohen,
  Evans-Searles,Searles-Evans, Searles2001, Gallavotti-Cohen, Kurchan,
  Lebowitz-Spohn, Crooks1998,Crooks-1,Crooks-2}, which accounts for the
measure of the likelihood of trajectories violating the second law of
thermodynamics.

For a small-scale heat engine connected to two heat reservoirs, the
efficiency becomes a fluctuating quantity, whose value changes from
one measurement to the other. Hence, it is described by a probability
distribution $P(\eta,t)$. In particular, one is interested in its
large deviation form \cite{Touchette} $P(\eta,t)\sim e^{t J(\eta)}$,
where $J(\eta)$ is the large deviation function defined as
$J(\eta)=\lim_{t \to \infty}\frac{1}{t}\ln P(\eta)$. It captures the
large time statistics of the efficiency of the stochastic engine.  In
a recent study, Verley \emph{et al.} \cite{verley-1} computed the large
deviation function $J(\eta)$ using FT for microscopic heat engine
using two set of examples: work to work converter engine and a
photoelectric device. They have shown that the Carnot efficiency is
least likely in the long time limit, which is a remarkable
result. Moreover, the large deviation function has two extrema: a
maximum corresponds to the most probable efficiency, and the minimum
occurs at Carnot efficiency.  In a similar context, Verley \emph{et
al.} \cite{verley-2}, found an efficient way to compute the large
deviation function of stochastic efficiency using the cumulant
generating function of entropy productions for a small engine with
finite state space. This method was verified by considering an example
of a stochastic engine made up of a system of two states where each of
these states is coupled to a heat reservoir at a distinct
temperature. To drive this system in the nonequilibrium state, a
time-dependent periodic field is applied. They have computed the large
deviation function for stochastic efficiency which supported the
prediction given in Ref. \cite{verley-1}.  Gingrich \emph{et al.} \cite{Gingrich}
computed the finite time probability density function for stochastic
efficiency of a two-level heat engine using time-asymmetric driving in
a cyclic process.  Polettini \emph{et al.} \cite{Polettini} derived the
probability density function for stochastic efficiency where
thermodynamic fluxes are distributed by a multivariate Gaussian
distribution. Using FT for entropy production, it is shown that
the probability of efficiency larger than the Carnot one, called
\emph{super-Carnot efficiency}, is favored by trajectories violating
the second law of thermodynamics. Moreover, the distribution function has
two maxima and one minimum: one maximum corresponds to the most probable
efficiency, while the other is at efficiency larger than the Carnot
efficiency. The location of the minimum is at the Carnot
efficiency. It is observed that the other maximum does not appear in
the large deviation function because in the long time limit that
maximum occurs at infinity.  Proesmans \emph{et al.} \cite{Proesmans-1}
considered an effusion process using two compartments at different
temperatures and chemical potentials, where particles flow from a
compartment at a higher temperature and low chemical potential to the
compartment at a low temperature and high chemical potential. In the
finite and long time limit, the distribution for the stochastic
efficiency is computed for this effusion engine. Some of these models
are briefly discussed in Ref. \cite{Proesmans-2}.  In the case of
isothermal energy transformation \cite{Proesmans-3}, authors
considered a Brownian particle in a harmonic potential driven by a duo
of time-periodic forces. This setup is used as an engine which
converted the Gaussian stochastic input work to Gaussian stochastic
output work. They have reproduced the latest discovered connection
between different operational regimes (maximum power, maximum
efficiency, minimum dissipation)
\cite{Benenti,Naoto,Proesmans-4}. Moreover, the probability density
function for stochastic efficiency is also computed, and all of these
results were verified experimentally by them.  Park \emph{et al.} \cite{Park}
modeled an engine which is driven by time-independent (time-symmetric)
driving. In contrast to Refs. \cite{verley-1,verley-2}, the phase space is
found to be continuous with infinite microstates, and it has been shown that the
large deviation function does not follow the universal nature as
mentioned in Refs. \cite{verley-1,verley-2}.

In this paper, we mainly focus on an isothermal energy converter where
a system consists of a Brownian particle coupled to a heat bath at a
constant temperature. In the absence of external forces, total entropy
production is identically zero as the system, described by the
velocity variable, enjoys equilibrium. The given system is maintained
in nonequilibrium steady state using two time-dependent stochastic
Gaussian external forces. This system functions as an engine which
converts one form of the work (\emph{input work}) into another form
(\emph{output work}). Note that this engine is different from the usual heat engines
where the working substance undergoes the cyclic transformation between two temperatures. 
Such an isothermal engine can be seen in biological systems,  
for example, adenosine triphosphate functions as an energy converter in the cell \cite{Bruce,Sugawa}.
The work done by these forces is stochastic
random variables. The efficiency of these isothermal engines is
defined by the ratio of the work done against the load force to the
work done by the drive force, which is also a stochastic quantity. We
compute the distribution of stochastic efficiency from the joint
distribution of work done against the load force (output work) and
work done by the drive force (input work). There are three important
features of this paper: (1) We have applied stochastic forces to drive
the system out of equilibrium, (2) FT for the joint probability
distribution of input and output work does not remain valid for all
strength of stochastic forces, and (3) the phase space is continuous with
infinite microstates. While the first two features were not introduced in
this context earlier as reported in Refs. \cite{Proesmans-3,Park}, the third
feature is similar to as mentioned in Ref. \cite{Park}.

The remainder of the paper is organized as follows. In \sref{model},
we give a model system of an engine which converts the input work to
the output work. \ssref{FP} contains the calculation of the joint
characteristic function of the input and output work,
$Z(\lambda_1,\lambda_2)\sim g(\lambda_1,\lambda_2) e^{(t/t_\gamma)
  \mu(\lambda_1,\lambda_2)}$ at large $t$. In \sref{pdf-w1w2}, we
discuss the method to invert the characteristic function
$Z(\lambda_1,\lambda_2)$ to get the probability density function
$P_t(W_1,W_2)$. In \sref{glambda}, we analyze the singularity present
in $g(\lambda_1,\lambda_2)$. In \sref{no-contour}, we write the
asymptotic expression for the joint probability density $P_t(W_1,W_2)$
using a saddle point approximation in the absence of a singularity in the
prefector $g(\lambda_1,\lambda_2)$, and in \sref{contour}, we
discuss the joint probability density function $P_t(W_1,W_2)$ in the
presence of a singularity in $g(\lambda_1,\lambda_2)$. FT for
$P_t(W_1,W_2)$ is discussed in \sref{FT}. In \sref{case1}, we give the
expression for the probability density function for stochastic
efficiency $P(\eta,t)$ when $g(\lambda_1,\lambda_2)$ does not have
singularities, and the result for this case is shown in
\sref{result-1}. In the case, when $g(\lambda_1,\lambda_2)$ has
singularities, we discuss the methodology to get the asymptotic
expression for $P(\eta,t)$ in \sref{case2}, and results in this case
are shown in \srefss{result-2} and \srefs{result-3}. We
summarized our paper in \sref{summ}. Some of the results are given in the
\aref{sing-contour}.

\section{Isothermal work-to-work converter engine}
\label{model}
Consider a Brownian particle of mass $m$ immersed in a heat bath of a
constant temperature $T$. In the absence of the external driving to
the particle, this system reaches an equilibrium state as given by
the Gibbs-Boltzmann measure. To model this Brownian particle as an engine,
we apply two different time-dependent forces on it. These forces are
called load force and drive force. The function of drive force is to
drive Brownian particle against the load force. For simplicity, we
assume these forces are uncorrelated. The dynamics of the engine is
governed by the Langevin equation,
\begin{equation}
m\dot{v}=-\gamma v+\xi(t)+f_{1}(t)+f_2(t),
\label{dynamics}
\end{equation} 
where $v$ is the velocity of the Brownian particle, $\gamma$ is the
dissipation constant, and $\xi(t)$ is the Gaussian white noise from
the bath, with mean zero and variance $\langle
\xi(t)\xi(t^\prime)\rangle=2 T\gamma \delta(t-t^\prime)$, according to
the fluctuation-dissipation theorem.  We set Boltzmann's constant to
unity throughout the calculation.  The load force $f_{1}(t)$ and the
drive force $f_{2}(t)$ are external stochastic Gaussian forces with
mean zero and variances $\langle f_i(t)f_j(t^\prime)\rangle=
\delta_{i,j}\bar{f}_i^2 \delta(t-t^\prime)$. They are uncorrelated
with the thermal noise $\langle f_i(t)\xi(t^\prime)\rangle=0$ for all
$t,t^\prime$. It turns out that only the relative strengths
amongst the external forces and the thermal noise are important, not their absolute values.  Therefore, we set $\bar{f}_1^2= 2 T\gamma
\theta$ and $\bar{f}_2^2= 2 T\gamma \theta \alpha^2$, where $\theta$
and $\alpha$ are positive parameters.

Multiplying both sides of \eref{dynamics} by $v$, and integrating with
respect to time from $0$ to $t$, yields the conservation of energy
relation (\emph{first law of thermodynamics})
\begin{equation}
\Delta E=Q+W_1+W_2,
\end{equation}
where 
\begin{align}
\Delta E&=\frac{m}{2T}[v^2(t)-v^2(0)],&\\ Q&=\dfrac{1}{T}\int_0^t
dt'\ [\xi(t^\prime)-\gamma v(t^\prime)]v(t^\prime),\label{Q-int}
\\ W_1&=\dfrac{1}{T}\int_0^t dt'\ f_1(t^\prime)v(t^\prime),\label{W1}
\\ W_2&=\dfrac{1}{T}\int_0^t dt'\ f_2(t^\prime)v(t^\prime).\label{W2}.
\end{align}
Here, we measure change in the internal energy $\Delta E$, heat
absorbed from the surrounding bath $Q$, and work done by load and drive
forces $W_1$ and $W_2$, respectively, in the scale of temperature of
the heat bath. The integrals given in \erefss{Q-int}--\erefs{W2} 
follow the Stratonovich rule of integration.

It is clear from \eref{dynamics} that the velocity $v$ depends
linearly on both thermal noise $\xi(t)$ and external Gaussian forces
$f_1(t)$ and $f_2(t)$. Therefore, the distribution of $v(t)$ is
Gaussian, where the mean and the variance can easily be computed from
\eref{dynamics}. In the limit $t\to\infty$, the mean velocity becomes
zero, and the variance is given by
\begin{equation}
\left[\langle v^2(t)\rangle-\langle v(t)\rangle^2\right]_{t\to\infty}
=\dfrac{T(1+\theta+\theta \alpha^2)}{m}.
\end{equation}
On the other hand, $W_1$ and $W_2$ given in \erefss{W1}
   and \erefs{W2}, respectively, depend on thermal noise $\xi(t)$ and
   external Gaussian forces $f_1(t)$ and $f_2(t)$
   quadratically. Thus, the joint distribution $P_t(W_1,W_2)$ is not
 expected to be Gaussian.

The quantity of interest is the the efficiency of a stochastic engine
$\eta$ which converts the input work $W_2$ to the output work $-W_1$:
\begin{equation}
\eta=-\dfrac{W_1}{W_2}.
\end{equation}
 The distribution of this stochastic efficiency
$P(\eta,t)$ is computed from the joint distribution of input and
output work $P_t(W_1,W_2)$ by integrating over $W_1$ while using the Dirac
delta function $\delta(\eta+W_1/W_2)$.  Therefore,
\begin{align}
P(\eta,t)&=\int_{-\infty}^\infty dW_2\ |W_2|\ P_t(-\eta W_2,W_2),
\label{peta-int}
\end{align}
where $|W_2|$ is the Jacobian.

Note that, when the joint distribution $P_t(W_1,W_2)$ is Gaussian
(that is not the case here),
\begin{equation}
P_t(W_1=w_1 t,W_2=w_2 t)=\dfrac{1}{t \sqrt{(2 \pi)^2\det{C}}}e^{-\frac{t}{2} \bar{w}^T C^{-1} \bar{w}},
\end{equation}
using \eref{peta-int}, the distribution of the stochastic efficiency
$P(\eta, t)$ can easily be shown to be~\cite{verley-1}
  \begin{align}
&P(\eta,t)=\dfrac{e^{j(\eta) t}}{\sqrt{(2\pi)^2\det C}}\nonumber\\&\times\dfrac{[2 e^{-\frac{t}{2} a(\eta) b(\eta)^2}+b(\eta) \sqrt{2 \pi t\ a(\eta)} \erf\big(b(\eta)\sqrt{a(\eta)\ t/2}\big)]}{a(\eta)},
\label{p-eta-gaussian}
\end{align}
where $\bar{w}^T=(w_1-\mu_1,w_2-\mu_2)$, $ C_{ij}=(\langle W_i
W_j\rangle-\langle W_i\rangle\langle W_j\rangle)/t$, $\mu_i=\langle
W_i\rangle/t$, and
\begin{align}
j(\eta)&=-\dfrac{1}{2}\dfrac{(\eta \mu_2+\mu_1 )^2}{C_{22} \eta^2+2C_{12}\eta+C_{11}}\\
a(\eta)&=\dfrac{(\eta C_{22}+C_{12})^2+\det{C}}{C_{22} \det{C}}\\
b(\eta)&=\dfrac{(C_{11}+C_{12}\eta)\mu_2-(C_{12}+C_{22}\eta)\mu_1}{C_{22} \eta^2+2C_{12}\eta+C_{11}}.
\end{align}
In \eref{p-eta-gaussian}, $\erf(u)$ is the error function given by
\begin{equation}
\erf(u)=\dfrac{2}{\sqrt{\pi}}\int_0^u  e^{-x^2}\ dx.
\label{error-function}
\end{equation}

The goal of this paper is to understand the statistics of the
efficiency fluctuation when $P_t(W_1,W_2)$ is non-Gaussian.

\section{Fokker-Planck equation}
\label{FP}

To compute $P_t(W_1,W_2)$, it is convenient to first compute the
characteristic function $Z(\lambda_1,\lambda_2)=\langle\exp(-\lambda_1
W_1-\lambda_2 W_2)\rangle$. The conditional characteristic function
$Z(\lambda_1,\lambda_2,v,t|v_0)$ for fixed initial and final
conditions, $v(0)=v_0$ and $v(t)=v$, satisfies
\begin{equation}
\dfrac{\partial Z(\lambda_1,\lambda_2,v,t|v_0)}{\partial t}=\mathcal{L}_{\lambda_1,\lambda_2}Z(\lambda_1,\lambda_2,v,t|v_0),
\label{FP-Z-eqn}
\end{equation} 
where the differential operator $\mathcal{L}_{\lambda_1,\lambda_2}$ is
given by
\begin{align}
\mathcal{L}_{\lambda_1,\lambda_2}=&\bigg[\dfrac{T \gamma(1+\theta+\theta \alpha^2)}{m^2}\dfrac{\partial^2 }{\partial v^2}
+\dfrac{\gamma[1+2 \theta(\lambda_1+\alpha^2\lambda_2)]}{m} v\dfrac{\partial}{\partial v}\nonumber\\
&+\bigg\{\dfrac{\gamma[1+\theta(\lambda_1+\alpha^2\lambda_2)]}{m}
+\dfrac{\lambda_1^2+\alpha^2\lambda_2^2 }{T}\gamma \theta v^2\bigg\}\bigg]. 
\end{align}
The differential equation given in \eref{FP-Z-eqn} is subject to
initial condition $Z(\lambda_1,\lambda_2,v,0|v_0)=\delta(v-v_0)$. Note
that, putting $\lambda_1=\lambda_2=0$ in
$Z(\lambda_1,\lambda_2,v,t|v_0)$, gives the distribution of velocity
$v$ at time $t$ for given initial velocity $v_0$,
$P(v,t|v_0)=Z(0,0,v,t|v_0)$. Consequently, the steady-state velocity
distribution is given by $P_{ss}(v)=Z(0,0,v,t\to\infty|v_0)$,
independent of $v_0$.

The characteristic function $Z(\lambda_1,\lambda_2)$ is obtained from
$Z(\lambda_1,\lambda_2,v,t|v_0)$ by averaging over the initial
velocity with respect to the steady-state distribution $P_{ss}(v_0)$
and integrating over the final velocity $v$: 
\begin{equation}
Z(\lambda_1,\lambda_2)=\int_{-\infty}^{+\infty} dv\ \int_{-\infty}^{+\infty} dv_0\ P_{ss}(v_0)\ 
Z(\lambda_1,\lambda_2,v,t|v_0). 
\label{ch-fn}
\end{equation}
To solve differential equation given in \eref{FP-Z-eqn}, we write
\begin{equation}
Z(\lambda_1,\lambda_2,v,t|v_0)=e^{-\frac{1}{2T}[U(v)-U(v_0)]}\ \psi_{\lambda_1,\lambda_2}(v,t|v_0). 
\label{mapping-Z}
\end{equation}
 It follows that, for the particular choice
\begin{equation}
U(v)=\dfrac{m[1+2\theta(\lambda_1+\alpha^2\lambda_2)]}{2(1+\theta+\theta
  \alpha^2)}  v^2,
\end{equation}
$\Psi_{\lambda_1, \lambda_2}(v,t|v_0)$ satisfies the Schr\"odinger
equation in the imaginary time $-i\hbar t$ (and identifying
$[T\gamma(1+\theta+\theta \alpha^2)/m^2]$ with $[\hbar^2/(2m_q)]$ in
the quantum problem),
\begin{align}
\dfrac{\partial \psi_{\lambda_1,\lambda_2}(v,t|v_0)}{\partial t}=\bigg[\dfrac{T \gamma(1+\theta+\theta \alpha^2)}{m^2}\dfrac{\partial^2}{\partial v^2}- V(v)\bigg] \psi_{\lambda_1,\lambda_2}(v,t|v_0),\nonumber\\
\label{psi-eqn}
\end{align}
for a quantum harmonic oscillator (QHO), where 
\begin{equation}
V(v)=\frac{1}{2} m_q w_q^2 v^2 - \dfrac{\gamma}{2m},
\end{equation}
with the identification
\begin{equation}
 m_q w_q^2 =\dfrac{\gamma}{T}\bigg[\dfrac{[1+2\theta(\lambda_1+\alpha^2\lambda_2)]^2}{2(1+\theta+\theta \alpha^2)}-2\theta(\lambda_1^2+\alpha^2\lambda_2^2) \bigg].
\end{equation}
Thus, $\psi_{\lambda_1,\lambda_2}(v,t|v_0)=\langle v|e^{-\hat{H}t}|v_0
\rangle$ is recognized as the propagator of the QHO, which is known
exactly. For our purpose, it is convenient to expand
  $\psi_{\lambda_1,\lambda_2}(v,t|v_0)$ in the eigenbasis
  $\{\psi_n(v)\}$ of $\hat{H}$ as
\begin{align}
  \psi_{\lambda_1,\lambda_2}(v,t|v_0)=\sum_{n=0}^{\infty}e^{-t
      E_n(\lambda_1,\lambda_2)}
  \psi_n(v)\psi_n^*(v_0),
  \label{psi-expansion}
\end{align}
where the eigenvalues are given by
\begin{equation}
E_n=\left(n+\frac{1}{2}\right) \hbar w_q - \dfrac{\gamma}{2m},\quad
n=0,1,2,\dotsc.
\end{equation}
From the above identification between the quantum and the stochastic
problem, we have
\begin{equation}
\hbar w_q = (\gamma/m) \nu(\lambda_1,\lambda_2)
  \end{equation}
with
\begin{align}
  \nu(\lambda_1,\lambda_2)=\big[1+4\theta\{\lambda_1(1-\lambda_1)+\alpha^2\lambda_2(1-\lambda_2)\nonumber\\ -\alpha^2\theta(\lambda_1-\lambda_2)^2\}\big]^{1/2}.
\label{nu-lambda}
\end{align}

In the long time limit, \eref{psi-expansion} is dominated by the $n=0$
(ground state) term.  Thus, for large $t$, \eref{mapping-Z} becomes
\begin{equation}
Z(\lambda_1,\lambda_2,v,t|v_0) =
e^{(t/t_\gamma)\ \mu(\lambda_1,\lambda_2)}
\Psi(v,\lambda_1,\lambda_2)\chi(v_0,\lambda_1,\lambda_2) +\dotsb
\end{equation}
where 
$t_\gamma=m/\gamma$ is the viscous
  relaxation time, and 
  \begin{align}
    \label{mu-lambda}
  \mu(\lambda_1,\lambda_2)&=\frac{1}{2}[1-\nu(\lambda_1,\lambda_2)],\\
  \Psi(v,\lambda_1,\lambda_2) & = A_0 \,e^{-\frac{\beta}{2}U(v)} \psi_0(v),\\
  \chi(v_0,\lambda_1,\lambda_2) & = A_0^{-1}\, e^{\frac{\beta}{2}U(v_0)} \psi_0^*(v_0),
\end{align}
where $A_0$ is an arbitrary function of $\lambda_1$ and $\lambda_2$.
Note that $\chi(v_0,\lambda_1,\lambda_2)$ and
$\Psi(v,\lambda_1,\lambda_2)$, respectively are also the left and right
eigenfunctions of the differential operator
$\mathcal{L}_{\lambda_1,\lambda_2}$ corresponding to the largest
eigenvalue $\mu(\lambda_1,\lambda_2)$.  Using the ground state
eigenfunction of the QHO, with a particular choice of $A_0$, it can
easily be found that
 \begin{align}
\Psi(v,\lambda_1,\lambda_2)&=\sqrt{\dfrac{m\gamma\ \nu(\lambda_1,\lambda_2)}{2\pi\ \Theta}}\nonumber\\ &\times
\exp\bigg(-\frac{m\gamma}{4\Theta}[\nu(\lambda_1,\lambda_2)+1+2\theta(\lambda_1+\alpha^2\lambda_2)
]v^2\bigg),\\ \chi(v_0,\lambda_1,\lambda_2)&=\exp\bigg(-\frac{m\gamma}{4\Theta}[\nu(\lambda_1,\lambda_2)-1-2\theta(\lambda_1+\alpha^2\lambda_2)]
v_0^2\bigg),
\end{align}
with $\Theta=T\gamma (1+\theta+\theta \alpha^2)$.  The left and right
eigenfunctions satisfy the normalization condition
\begin{equation}
\int_{-\infty}^{+\infty}\chi(v,\lambda_1,\lambda_2) \Psi(v,\lambda_1,\lambda_2)\ dv=1.
\end{equation} 
From the above expressions, we find that $\mu(0,0)=0$ and
$\chi(v_0,0,0)=1$. Therefore, the steady state distribution
$P_{ss}(v)=Z(0,0,v,t\to\infty|v_0)$ of the velocity is given by
\begin{align}
P_{ss}(v)=\Psi(v,0,0)=\sqrt{\dfrac{m\gamma}{2\pi \Theta}}\exp\bigg[-\dfrac{m\gamma v^2}{2\Theta}\bigg].
\end{align}
The characteristic function $Z(\lambda_1,\lambda_2)$ is obtained after
carrying out integrals given in \eref{ch-fn},
\begin{equation}
Z(\lambda_1,\lambda_2) =  g(\lambda_1,\lambda_2) \exp[(t/t_\gamma)
  \mu(\lambda_1,\lambda_2)] + \dotsb.
\label{ch-fn-final}
\end{equation}
Here, the prefactor
\begin{equation}
g(\lambda_1,\lambda_2)=\dfrac{2\sqrt{\nu(\lambda_1,\lambda_2)}}{\sqrt{f^+(\lambda_1,\lambda_2)}\sqrt{ f^-(\lambda_1,\lambda_2)}},
\label{g-lambda}
\end{equation}
in which
$f^\pm(\lambda_1,\lambda_2)=1\pm2\theta(\lambda_1+\alpha^2\lambda_2)+\nu(\lambda_1,\lambda_2)$.
The first factor in the denominator of $g(\lambda_1,\lambda_2)$ is due
to the integration over the final velocity $v$, and the second factor
in the denominator of $g(\lambda_1,\lambda_2)$ comes from averaging
over the initial velocity $v_0$ with respect to steady-state
distribution $P_{ss}(v_0)$.

Note from \erefss{mu-lambda} and \erefs{g-lambda} that largest
eigenvalue $\mu(\lambda_1,\lambda_2)$ satisfies the Gallavotti-Cohen symmetry whereas the prefactor $g(\lambda_1,\lambda_2)$ does not,
\emph{i.e.,} $\mu(\lambda_1,\lambda_2)=\mu(1-\lambda_1,1-\lambda_2)$
and $g(\lambda_1,\lambda_2)\neq g(1-\lambda_1,1-\lambda_2)$.


\section{Joint Probability density function $P_t(W_1,W_2)$}
\label{pdf-w1w2}
\begin{figure}
\includegraphics[width=1.0\linewidth]{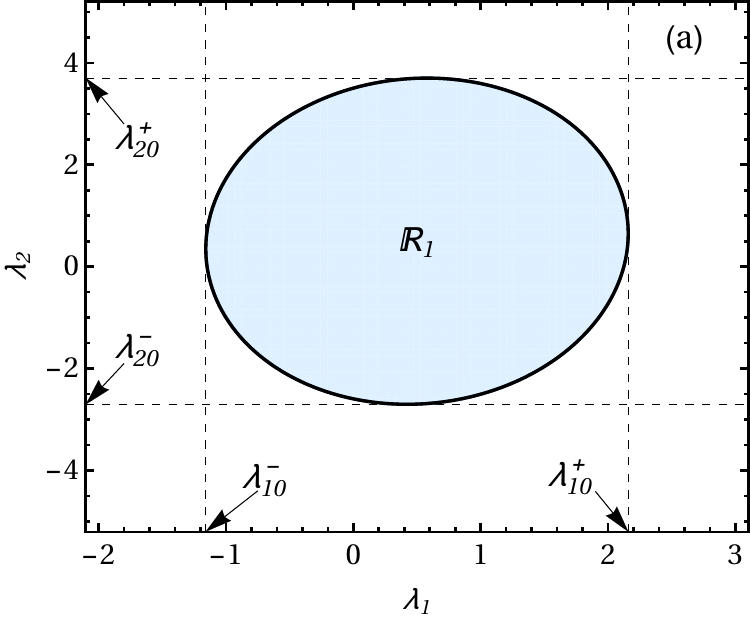}
\includegraphics[width=1.0\linewidth]{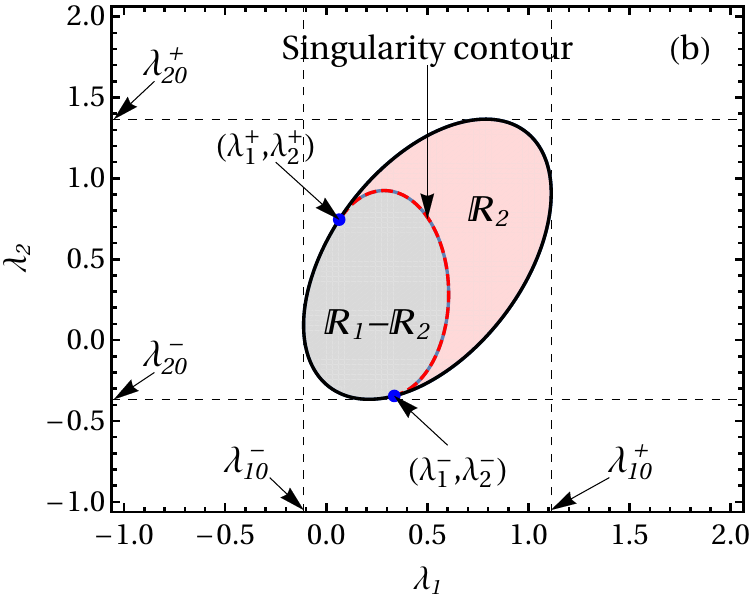}
\caption{Two scenarios are shown here. (a) $g(\lambda_1,\lambda_2)$
  is analytic for all $(\lambda_1,\lambda_2)\in \mathbb{R}_1$. (b) The light red region $\mathbb{R}_2$ represents the area where
  $g(\lambda_1,\lambda_2)$ is imaginary, whereas it is real in
  $\mathbb{R}_1-\mathbb{R}_2$. The red contour (thick dashed) is the singularity line and
  corresponds to \eref{red-contour}. End points $(\lambda_{1,2}^\pm)$
  are given in the \aref{sing-contour}.  In both cases, the black solid contour
  represents the region $\mathbb{R}_1$ bounded by
  $(\lambda_1(\phi),\lambda_2(\phi))$, $\phi\in[-\pi,\pi]$. Black
  dashed lines show the maximum and minimum values of
  $\lambda_{1}(\phi)$ and $\lambda_2(\phi)$ as given by
  \erefss{max-min-1} and \erefs{max-min-2}.}
\label{contours}
\end{figure}

The joint distribution of input and output work
  $P_t(W_1,W_2)$ can be obtained by inverting the characteristic
  function $Z(\lambda_1,\lambda_2)$ given in \eref{ch-fn-final}:
  \begin{equation}
P_t(W_1,W_2)=\int_{-i\infty}^{i\infty} \dfrac{d\lambda_1}{2 \pi i}\int_{-i\infty}^{i\infty} \dfrac{d\lambda_2}{2 \pi i}\ Z(\lambda_1,\lambda_2)\ e^{\lambda_1 W_1+\lambda_2 W_2}.
\end{equation}
Thus, for large $t$,
\begin{align}
P_t(W_1,W_2)&=\int_{-i\infty}^{i\infty} \dfrac{d\lambda_1}{2 \pi
  i}\int_{-i\infty}^{i\infty} \dfrac{d\lambda_2}{2 \pi
  i}\ g(\lambda_1,\lambda_2)\ e^{(t/t_\gamma)\ I_{w_1,w_2}(\lambda_1,\lambda_2)}\notag\\
&+\dotsb,
\label{pw_1w_2-int}
\end{align}
where $w_1=W_1 t_\gamma/t$ and $w_2=W_2 t_\gamma/t$ are scaled
variables. Here, the contours of integration are taken along
$\mathrm{Im}(\lambda_1)$ and $\mathrm{Im}(\lambda_2)$ axes passing
through the origin of the complex $(\lambda_1,\lambda_2)$ plane.
The function  $I_{w_1,w_2}(\lambda_1,\lambda_2)$ is given as
\begin{equation}
I_{w_1,w_2}(\lambda_1,\lambda_2)=\mu(\lambda_1,\lambda_2)+\lambda_1 w_1+\lambda_2 w_2.
\label{Iw_1w_2}
\end{equation}
It can be seen $\nu(\lambda_1,\lambda_2)$ is a
real and positive quantity when $(\lambda_1,\lambda_2)\in
\mathbb{R}_1$ where $\mathbb{R}_1$ is the region shown in \fref{contours}, bounded by $(\lambda_1(\phi),\lambda_2(\phi))$ in which
\begin{align}
&\lambda_1(\phi)=\dfrac{1}{2}\bigg[1+\sqrt{\dfrac{\alpha^2\theta}{1+\alpha^2\theta}}\sin\phi+\sqrt{\dfrac{1+\theta+\theta\alpha^2}{\theta(1+\alpha^2\theta)}}\cos\phi\bigg],\\
&\lambda_2(\phi)=\dfrac{1}{2}\bigg[1+\sqrt{\dfrac{1+\alpha^2\theta}{\alpha^2\theta}}\sin\phi\bigg], \quad\text{with}\quad \phi\in[-\pi,\pi]. \nonumber\\
\label{lambda12phi}
\end{align} 
Here, $(\lambda_1(\phi)$, $\lambda_2(\phi))$ is the parametric
representation of equation of ellipse [see \eref{nu-lambda}]
\begin{equation}
1+4\theta[\lambda_1(1-\lambda_1)+\alpha^2\lambda_2(1-\lambda_2)-\alpha^2\theta(\lambda_1-\lambda_2)^2]=0.
\end{equation}
The maximum and minimum values of  $\lambda_1(\phi)$ and $\lambda_2(\phi)$ (see black dashed lines in \fref{contours}) are
\begin{align}
\lambda_{10}^{\pm}&=\dfrac{1}{2}\bigg[1\pm\sqrt{1+\dfrac{1}{\theta}}\bigg]\label{max-min-1},\\
\lambda_{20}^{\pm}&=\dfrac{1}{2}\bigg[1\pm\sqrt{1+\dfrac{1}{\alpha^2\theta}}\bigg]\label{max-min-2},
\end{align}
where $+$ and $-$ signs correspond to maximum and minimum value, respectively. Consequently, $I_{w_1,w_2}(\lambda_1,\lambda_2)$ is also a real quantity when $(\lambda_1,\lambda_2)\in \mathbb{R}_1$.

The long-time result of the integral given in \eref{pw_1w_2-int} can be approximated using the saddle-point method. The saddle point ($\lambda_1^*,\lambda_2^*$) can be obtained by solving the following equations simultaneously:
\begin{align}
\dfrac{\partial I_{w_1,w_2}(\lambda_1,\lambda_2)}{\partial \lambda_1 }\bigg|_{\lambda_{1,2}=\lambda_{1,2}^*}&=0, \\ \dfrac{\partial I_{w_1,w_2}(\lambda_1,\lambda_2)}{\partial \lambda_2 }\bigg|_{\lambda_{1,2}=\lambda_{1,2}^*}&=0.
\label{saddle-points}
\end{align}
This gives
\begin{align}
\lambda_1^*(w_1,w_2)&=\dfrac{1}{2}\bigg[1-\dfrac{\alpha[w_1+(w_1+w_2)\theta]}{\Lambda}\bigg],\\
\lambda_2^*(w_1,w_2)&=\dfrac{-w_2-(w_1+w_2)\alpha^2\theta+\alpha \Lambda}{2 \alpha \Lambda },
\end{align}
where\\
$\Lambda=\sqrt{\theta [w_1^2 \alpha^2+w_2^2+(w_1+w_2)^2 \alpha^2 \theta+\alpha^2\theta(1+\theta+\theta \alpha^2)]}$.
Clearly, one can see that $(\lambda_1^*,\lambda_2^*)\in \mathbb{R}_1$.
Moreover, at the saddle point, the function $I(w_1,w_2):=I_{w_1,w_2}(\lambda_1^*,\lambda_2^*)$ reads as
\begin{equation}
I(w_1,w_2)=\dfrac{1}{2}\bigg[1+w_1+w_2-\dfrac{\Lambda}{\alpha \theta}\bigg].
\label{LDF}
\end{equation}

Now, to solve the integral given in \eref{pw_1w_2-int}, we have to
analyze whether $g(\lambda_1,\lambda_2)$ is analytic when
$(\lambda_1,\lambda_2)\in \mathbb{R}_1$. If there is no singularity
present in $g(\lambda_1,\lambda_2)$ between the origin of
the $(\lambda_1,\lambda_2)$ plane and saddle point
($\lambda_1^*,\lambda_2^*$), one can deform the contours of
integration through the saddle point ($\lambda_1^*,\lambda_2^*$) and
carry out saddle-point integration to approximate the integral given in
\eref{pw_1w_2-int} \cite{Apal-1,Apal-2,Touchette}. However, if
$g(\lambda_1,\lambda_2)$ contains a singularity between the saddle point and
the origin of $(\lambda_1,\lambda_2)$ plane, then the saddle-point
approximation will not be valid. In the following subsections, we
consider both cases.
\begin{figure}
\includegraphics[width=1.0\linewidth]{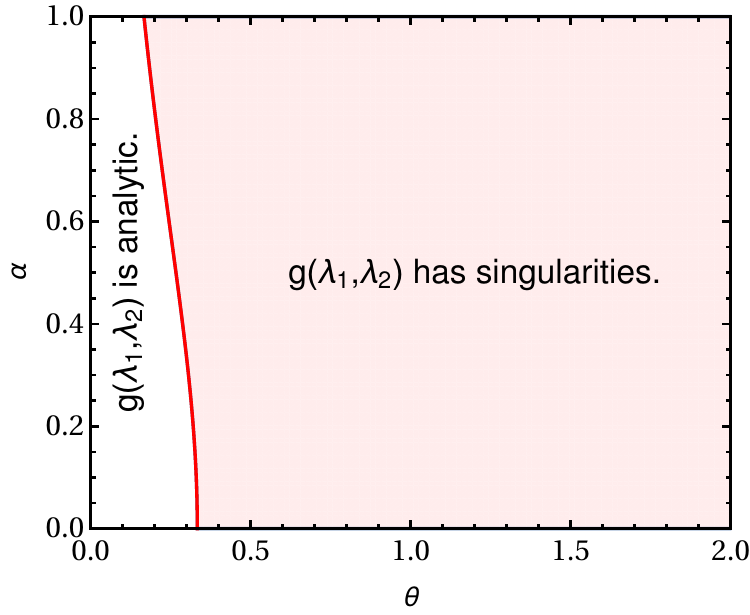}
\caption{Contour (red solid line) separates two regions depending upon if $g(\lambda_1,\lambda_2)$ has singularities or not.}
\label{phase-dig-1}
\end{figure}


\subsection{Analytic behavior of the correction term $g(\lambda_1,\lambda_2)$}
\label{glambda}
In $g(\lambda_1,\lambda_2)$, $f^+(\lambda_1,\lambda_2)>0$ for all
$\theta\in(0,\infty)$ and $\alpha\in(0,\infty)$ whereas the function
$f^-(\lambda_1,\lambda_2)$ can attain any sign depending on the values
of $\theta$ and $\alpha$.  It turns out that there can be two
scenarios, which are shown in \fref{contours}. In both scenarios,
$\mathbb{R}_1$ is the region bounded by contour
$(\lambda_1(\phi),\lambda_2(\phi))$ where $\phi\in[-\pi,\pi]$ (see black solid contour in \fref{contours}). In \fref{contours}(a),
$f^-(\lambda_1,\lambda_2)> 0$ for all $(\lambda_1,\lambda_2)\in
\mathbb{R}_1$, and hence $g(\lambda_1,\lambda_2)$ does not have any
singularity in the whole region $\mathbb{R}_1$. On the other hand, in
\fref{contours}(b), $f^-(\lambda_1,\lambda_2)\leq0$ in the region
$(\lambda_1,\lambda_2)\in \mathbb{R}_2$ and positive in
$(\lambda_1,\lambda_2) \in \mathbb{R}_1-\mathbb{R}_2 $. Hence
$g(\lambda_1, \lambda_2)$ has singularities given by the curve
\begin{equation}
f^-(\lambda_1,\lambda_2)=0.
\label{red-contour}
\end{equation}
We see two regions in the phase diagram in $(\alpha,\theta)$ plane shown in \fref{phase-dig-1}, which distinguish these two scenarios, and the equation of contour which separates these two regions is given by (see the \aref{sing-contour}) 
\begin{equation} 
(1+\alpha^2)\theta=1/3.
\end{equation}
In the next subsections, we will discuss both cases one by one.  


\subsection{Case 1: Singularity contour is absent}
\label{no-contour}
When there is no singularity contour present in the domain $\mathbb{R}_1$ [see \fref{contours}(a)], we can approximate the integral given in \eref{pw_1w_2-int} by the saddle-point method. Therefore, we get
\begin{equation}
P_t(W_1,W_2)\approx \dfrac{\tilde{g}(w_1,w_2)\ e^{(t/t_\gamma)\  I(w_1,w_2)}}{2 \pi\ t/t_\gamma \sqrt{|\tilde{H}(w_1,w_2)|}},
\label{pw_1w_2-saddle}
\end{equation}
where $\tilde{g}(w_1,w_2):=g(\lambda_1^*,\lambda_2^*)$, and
$\tilde{H}(w_1,w_2):= H(\lambda_1^*,\lambda_2^*)$ is the determinant
of the Hessian matrix,
\begin{align}
H(\lambda_1^*,\lambda_2^*)&=\bigg[\dfrac{\partial ^2 I(\lambda_1,\lambda_2)}{\partial \lambda_1^2}\dfrac{\partial ^2 I(\lambda_1,\lambda_2)}{\partial \lambda_2^2}\nonumber\\
&\quad\quad\quad\quad\quad\quad-\bigg(\dfrac{\partial ^2 I(\lambda_1,\lambda_2)}{\partial \lambda_1\partial \lambda_2}\bigg)^2\bigg]\bigg|_{\lambda_{1,2}=\lambda_{1,2}^*}\nonumber\\
&=\dfrac{4 \Lambda^4}{\alpha^2 \theta^2(1+\theta+\theta \alpha^2)^2}.
\label{Hessian-saddle}
\end{align}
The function $I(w_1,w_2)$ is given by \eref{LDF}. Here, $\tilde{H}(w_1,w_2)>0$
for all $\theta,\ \alpha,\ w_1$, and $w_2$ which implies that along axes
$\mathrm{Re}(\lambda_1)$ and $\mathrm{Re}(\lambda_2)$, function
$I_{w_1,w_2}(\lambda_1,\lambda_2)$ given in \eref{Iw_1w_2}, is minimum
at the saddle point $(\lambda_1^*,\lambda_2^*)$ . Therefore, contours
of integration are taken along the direction perpendicular to both
$\mathrm{Re}(\lambda_1)$ and $\mathrm{Re}(\lambda_2)$ axes of the
complex $(\lambda_1,\lambda_2)$ plane \cite{Apal-1,Apal-2}.

\subsection{Case 2: Singularity contour is present}
\label{contour}
When a singularity contour is present in the region
$(\lambda_1,\lambda_2)\in \mathbb{R}_1$ [see red contour (thick dashed) in
  \fref{contours}(b)], we have to compute the integral given in
\eref{pw_1w_2-int} carefully. In such a case, there will be two types of
contributions, namely, saddle and branch point
contributions. When the saddle point $(\lambda_1^*,\lambda_2^*)$ does
not cross the branch point contour given by \eref{red-contour} [see
  red contour (thick dashed) in \fref{contours}(b)] i.e., the saddle point does not
enter the light red region $\mathbb{R}_2$ of \fref{contours}(b),
the contribution is the same as given in \eref{pw_1w_2-saddle}. As the
saddle point crosses the branch point contour given by
\eref{red-contour}, then, the integral can not approximated with the
usual saddle-point solution, and one has to evaluate \eref{pw_1w_2-int}
carefully by taking into account of the
singularities~\cite{Apal-1,Apal-2,Lee,JDNoh}.

Since the equation of the singularity contour is
  given by \eref{red-contour}, in the $(w_1,w_2)$ plane, the contour
  separating these two regions (saddle and branch points) becomes
$h(w_1,w_2):=f^-(\lambda_1^*,\lambda_2^*)=0$. The joint
probability distribution of $W_1$ and $W_2$ is given as
\begin{equation}
P_t(W_1,W_2)=\begin{cases}
P_S(W_1,W_2,t) \quad\quad\quad \quad h(w_1,w_2)\gg0,\\
P_B(W_1,W_2,t) \quad\quad\quad \quad h(w_1,w_2)\ll0,\\
\end{cases}
\label{pw_1w_2-branch}
\end{equation}
where $P_S(W_1,W_2,t)$ and $P_B(W_1,W_2,t)$ are saddle and branch
point contributions, respectively. Signs $\ll$ and $\gg$ show that
both saddle and branch point contributions are valid away from the
singularity contour [see the red contour (thick dashed) in \fref{contours}(b)]
\cite{Apal-1,Apal-2}.

 
\subsection{Large deviation function and FT for joint distribution $P_t(W_1,W_2)$}
\label{FT}
The large deviation function is defined as
\begin{equation}
I(w_1,w_2)=\lim_{t/t_\gamma\ \to\ \infty}\frac{t_\gamma}{t}\ln{P_t(W_1,W_2)}, 
\end{equation}
and the large deviation form of joint distribution is usually written
as
\begin{equation}
P_{t}(W_1,W_2)\sim e^{(t/t_\gamma)I(w_1,w_2)}.
\end{equation}
For the distribution satisfying FT, it is seen that
\begin{equation}
\lim_{t/t_\gamma\ \to\ \infty}\dfrac{t_\gamma}{t}\ln{\bigg[\dfrac{P(W_1=+w_1 t/t_\gamma,W_2=+w_2 t/t_\gamma)}{P(W_1=-w_1 t/t_\gamma,W_2=-w_2 t/t_\gamma)}\bigg]}=w_1+w_2.
\end{equation}
When the above relation holds, the large deviation function satisfies a symmetry property given as
\begin{equation}
I(w_1,w_2)-I(-w_1,-w_2)=w_1+w_2 \quad \text{for all} \quad (w_1,w_2).
\label{symm-rel}
\end{equation}

The phase diagram given in \fref{phase-dig-1} characterizes regions
of analyticity for the prefactor $g(\lambda_1,\lambda_2)$. If
$g(\lambda_1,\lambda_2)$ does not have any singularity in the region
$(\lambda_1,\lambda_2)\in \mathbb{R}_1$, the dominant contribution to
the joint distribution $P_t(W_1,W_2)$ comes from the saddle-point
approximation as given by
\eref{pw_1w_2-saddle}. However, when the saddle point
  $(\lambda_1^*,\lambda_2^*)$ crosses the branch point contour shown
  in \fref{contours}(b) [see light red region $\mathbb{R}_2$ of
    \fref{contours}(b)], the contribution to $P_t(W_1,W_2)$ comes
  from both saddle and branch points as given by
  \eref{pw_1w_2-branch}.
  Thus, for the region where
$g(\lambda_1,\lambda_2)$ does not have any singularity (see
  \fref{phase-dig-1}), the large deviation function $I(w_1,w_2)$ is
given by \eref{LDF} and satisfies the relation given in \eref{symm-rel}, and
hence, the fluctuation theorem is satisfied. On the other hand, when
$g(\lambda_1,\lambda_2)$ has singularities, the fluctuation theorem would
not be satisfied for large $(w_1,w_2)$.

\section{Probability density function of Stochastic efficiency $P(\eta,t)$} 
\label{Peta}
After computing the asymptotic form of $P_t(W_1,W_2)$, we have to carry out one more integral given in \eref{peta-int} to get the probability density function of stochastic efficiency.
There are two cases, which we will discuss in the next subsections.


\subsection{Case 1: $g(\lambda_1,\lambda_2)$ does not have singularities}
\label{case1}
When the asymptotic form of $P_t(W_1,W_2)$ is given by only the saddle-point contribution [see \eref{pw_1w_2-saddle}], then the integral
given in \eref{peta-int} can also be computed using the saddle-point
method. In that case, by solving the saddle-point equation
\begin{equation}
\dfrac{\partial I(-\eta w_2,w_2)}{\partial w_2}\bigg|_{w_2=w_2^*}=0,
\end{equation}
we find the saddle point $w_2^*(\eta)$ as
\begin{equation}
w_2^*(\eta)=\dfrac{(1-\eta)\alpha^2\theta\sqrt{1+\theta+\theta
    \alpha^2}}{\sqrt{(1+\eta^2\alpha^2)[1+\eta^2\alpha^2+\alpha^2\theta(1-\eta)^2]}}. 
\label{w_2s}
\end{equation} 
Finally, the probability density function for stochastic efficiency is
given by
\begin{equation}
P(\eta,t)\approx\dfrac{t \tilde{g}(-\eta w_2^*,w_2^*)\zeta(\eta,t)}{2
  \pi t_\gamma\sqrt{\tilde{H}(-\eta w_2^*,w_2^*)}}  \exp[(t/t_\gamma)\ I(-\eta w_2^*,w_2^*)],
\label{eqn2}
\end{equation}
where
\begin{equation}
 \zeta(\eta,t)=\dfrac{e^{-K Y^2}+\sqrt{\pi K}Y\erf(\sqrt{K}Y)}{K},
\end{equation}
with $Y=w_2^*(\eta)$, $K=t/(2t_\gamma) \big|\partial^2 I(-\eta w_2,w_2)/\partial w_2^2\big|_{w_2^*}$,
and 
$\erf(u)$ is given by \eref{error-function}.

The large deviation function $J(\eta):=I(-\eta w_2^*,w_2^*)$ is given by
\begin{equation}
J(\eta)=\dfrac{1}{2}\bigg[1-\sqrt{\dfrac{(1+\alpha^2\eta^2)(1+\theta+\theta \alpha^2)}{1+\alpha^2[\eta^2+(1-\eta)^2\theta]}}\bigg]
\end{equation} 
The large deviation function  $J(\eta)$ for stochastic efficiency has two extrema. The minimum occurs at $\eta^*=1$ while the maximum is at $\bar{\eta}=-\alpha^{-2}$. The efficiency at which the large deviation function is minimum is called \emph{ an analog of the Carnot efficiency} \cite{Polettini} as this is essentially the maximum value that the efficiency of a reversible engine can achieve in macroscopic systems.
At the efficiency $\bar{\eta}$, $J(\bar{\eta})=J'(\eta=\bar{\eta})=0$, which are the properties of a large deviation function.

\subsection{Numerical simulation}
\label{result-1}
\begin{figure}[!h]
\includegraphics[width=1.0\linewidth]{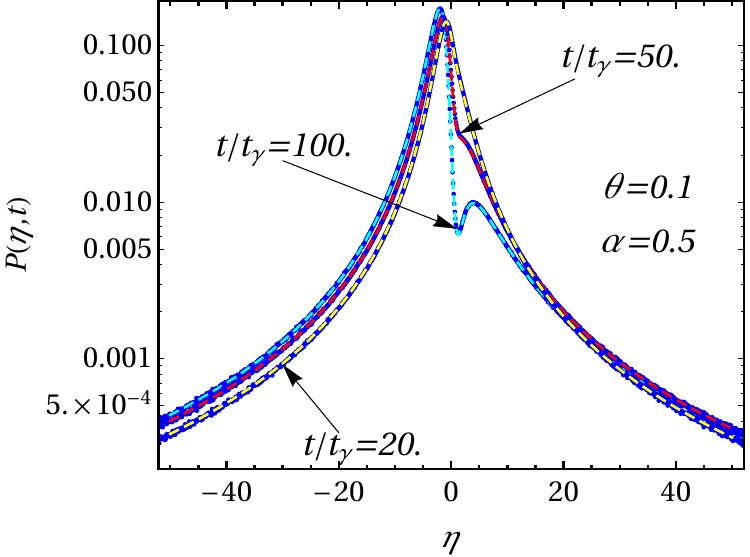}
\caption{Distribution of stochastic efficiency $P(\eta,t)$ is plotted against the stochastic efficiency $\eta$ for $\theta=0.1$ and $\alpha=0.5$ at three different times: $t/t_\gamma=20,50,100$.  Yellow $(t/t_\gamma=20)$, red $(t/t_\gamma=50)$, and cyan $(t/t_\gamma=100)$ dashed lines are plotted for analytical expression given in \eref{eqn2} while the blue dots are for the numerical simulations at the corresponding times $t/t_\gamma$. }
\label{dist-phase-1}
\end{figure}

We compare the analytical form given in \eref{eqn2} with the numerical simulation. We take parameter $\theta=0.1$ and $\alpha=0.5$ at three different times: $t/t_\gamma=20$, $t/t_\gamma=50$, and $t/t_\gamma=100$. \ffref{dist-phase-1}  shows a very good agreement with simulation and theoretical prediction. 


\subsection{Case 2: $g(\lambda_1,\lambda_2)$ has singularities}
\label{case2} 
 When $g(\lambda_1,\lambda_2)$ has singularities in the region
 $(\lambda_1,\lambda_2) \in \mathbb{R}_1$ [see \fref{contours}(b)], we need to be careful while
 computing the asymptotic form of $P_t(W_1,W_2)$, as given by
 \eref{pw_1w_2-branch}.

\begin{figure}[!h]
\includegraphics[width=1.0\linewidth]{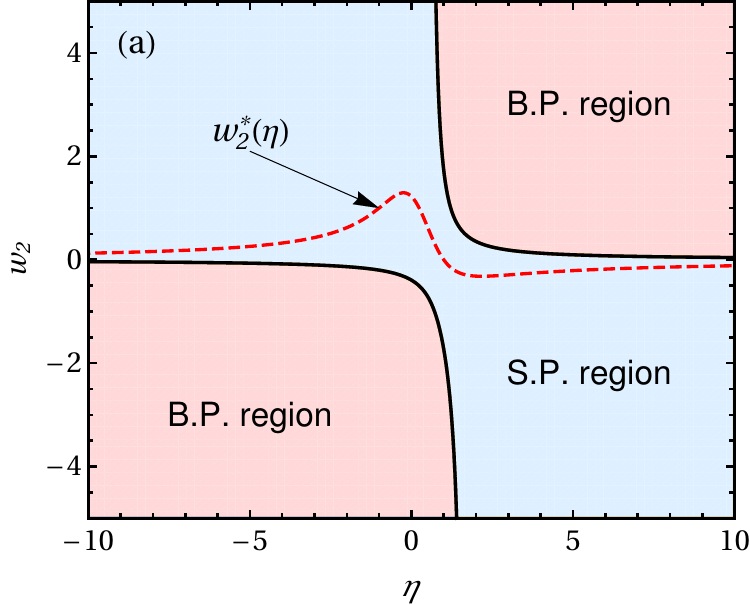}
\includegraphics[width=1.0\linewidth]{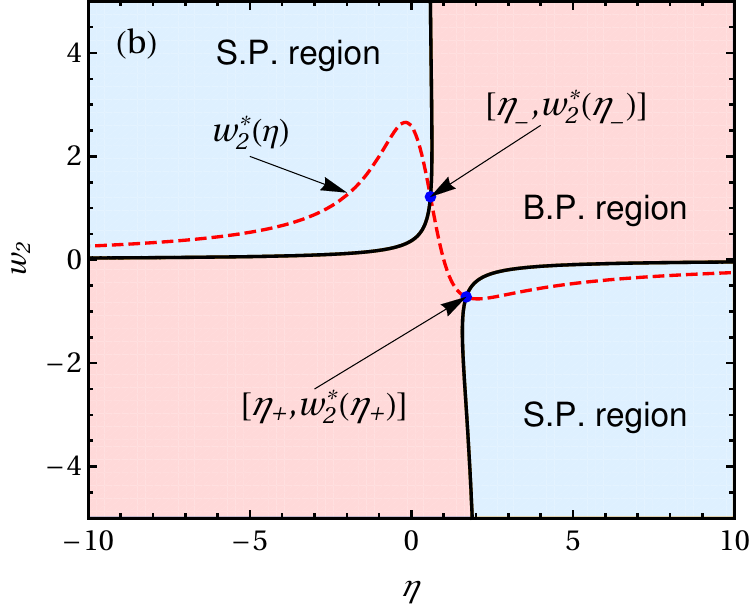}
\caption{Two possibilities are shown. (a) Saddle point $w_2^*(\eta)$ stays in the saddle-point region of $P_t(-\eta W_2,W_2)$. (b) Saddle point $w_2^*(\eta)$ stays in both the saddle- and branch-point regions of $P_t(-\eta W_2,W_2)$. Light blue (S.P. region) [$h(-\eta w_2,w_2)>0$] and light red (B.P. region) [$h(-\eta w_2,w_2)<0$] shaded areas are for saddle- and branch-point contributions of $P_t(-\eta W_2,W_2)$, respectively. Black solid line corresponds to $h(-\eta w_2,w_2)=0$. Blue points are given by $(\eta_-,w_2^*(\eta_-))$ and $(\eta_+,w_2^*(\eta_+))$.}
\label{saddle-eta-w2}
\end{figure}

It turns out that the saddle point $w_2^*(\eta)$ given in \eref{w_2s}, from the saddle-point contribution of $P_t(-\eta W_2,W_2)$ stays either in saddle point region (possibility I) or in both
saddle- and branch-point regions (possibility II) of $P_t(-\eta
w_2,w_2)$ depending upon parameters $\theta$ and $\alpha$ as shown in
\fref{saddle-eta-w2}. In \fref{saddle-eta-w2} light blue (S.P. region) and light red (B.P. region)
regions correspond to the saddle- and branch-point contributions of
joint distribution $P_t(-\eta W_2,W_2)$, respectively.

As saddle point $w_2^*(\eta)$ intersects the contour $h(-\eta
 w_2,w_2)=0$, it satisfies
\begin{equation}
h\big(-\eta_\pm w_2^*(\eta_\pm),w_2^*(\eta_\pm)\big)=0,
\end{equation}
in which
\begin{equation}
\eta_\pm=\dfrac{4\alpha\theta\pm\sqrt{3(1+\theta+\theta \alpha^2)}\sqrt{-3+\theta+\theta \alpha^2}}{\alpha[3(1+\theta)-\alpha^2\theta]}.
\end{equation}
Therefore, points where saddle point $w^*(\eta)$ intersects the
contour $h(-\eta w_2,w_2)=0$ are given by $(\eta_-,w_2^*(\eta_-))$ and
$(\eta_+,w_2^*(\eta_+))$.  The contour, which separates possibility I
from possibility II in the $(\theta, \alpha)$ space, is given by
the condition $\eta_+=\eta_-$, which results in
\begin{equation}
-3+\theta+\theta \alpha^2=0.
\end{equation}
It also follows from the fact that the efficiency is a real quantity,
and therefore, $\eta_\pm$ must be real, which implies
$(-3+\theta+\theta \alpha^2)\geq0$.

\begin{figure}
\includegraphics[width=1.0\linewidth]{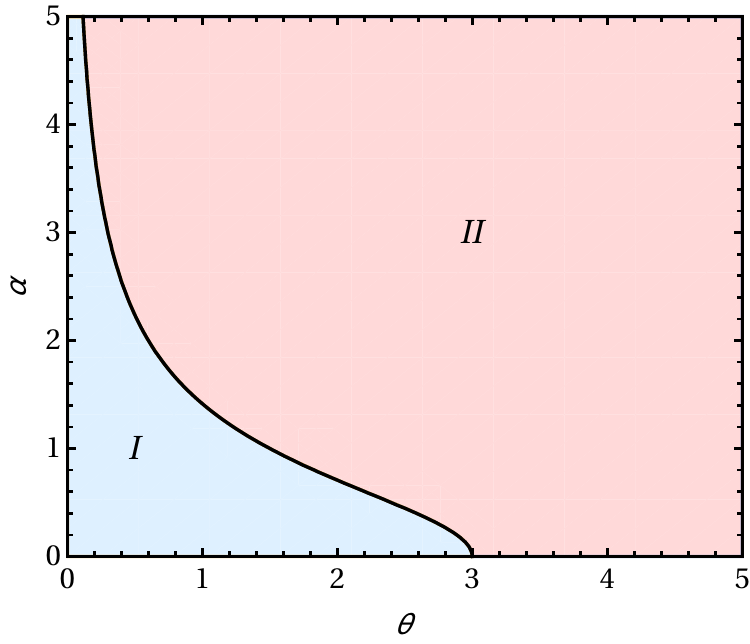}
\caption{Possibility I: The distribution of efficiency $P(\eta,t)$ can be computed by only saddle point solution given by \eref{pw_1w_2-saddle}. Possibility II: The distribution of efficiency $P(\eta,t)$ requires the branch point contribution of $P_t(-\eta W_2,W_2)$.}
\label{phase-dig-3}
\end{figure}

Using the above equation, we can draw a phase diagram in the $(\theta,\alpha)$
plane as shown in \fref{phase-dig-3}.
In possibility I, saddle point $w^*(\eta)$ does not intersect the
contour given by $h(-\eta w_2,,w_2)=0$, and stays in the saddle point
region of joint distribution $P_t(-\eta W_2,W_2)$. Therefore, only the
saddle-point contribution of $P(-\eta W_2,W_2)$ is required to compute
the asymptotic expression of $P(\eta,t)$.  But, for possibility II,
we actually need to compute the branch-point contribution to calculate
the asymptotic expression for $P(\eta,t)$.  Therefore, the
distribution of efficiency $P(\eta,t)$ is given as
\begin{equation}
P(\eta,t)\begin{cases}
=\text{\eref{eqn2}}  \quad \quad    h(-\eta w_2,w_2)>0\\
\neq\text{\eref{eqn2}}  \quad\quad    h(-\eta w_2,w_2)<0.\\
\end{cases}
\end{equation}
The analytical computation of the joint distribution for
possibility II is not very illuminating. Nevertheless, we can
perform numerical saddle-point integration to calculate
$P(\eta,t)$. This method is described in the following subsection.

\subsection{Numerical saddle-point integration}
\label{num-saddle}
We write the integral given in \eref{pw_1w_2-int} as
\begin{align}
P_t(-\eta W_2,W_2)=\int_{-i\infty}^{i\infty} \dfrac{d\lambda_1}{2 \pi i}\int_{-i\infty}^{i\infty} \dfrac{d\lambda_2}{2 \pi i}\ g_0(\lambda_1,\lambda_2)\nonumber\\\times e^{(t/t_\gamma)\ f(\lambda_1,\lambda_2,\eta,w_2,t)},
\label{pw_1w_2-num}
\end{align}
with 
\begin{equation}
f(\lambda_1,\lambda_2,\eta,w_2,t)=I_{-\eta w_2,w_2}(\lambda_1,\lambda_2)-\dfrac{t_\gamma}{2 t}\ln{f^-(\lambda_1,\lambda_2)}.
\end{equation}
Here $g_0(\lambda_1,\lambda_2)$ is the analytic part of $g(\lambda_1,\lambda_2)$, given as
\begin{equation}
g_0(\lambda_1,\lambda_2)=\dfrac{2\sqrt{\nu(\lambda_1,\lambda_2)}}{\sqrt{f^+(\lambda_1,\lambda_2)}}.
\label{g0lambda}
\end{equation}
Therefore, the saddle point  ($\lambda_1^*(\eta,w_2,t),\lambda_2^*(\eta,w_2,t)$) is the solution of following equations simultaneously:
\begin{align}
\dfrac{\partial f(\lambda_1,\lambda_2,\eta,w_2,t)}{\partial \lambda_1}\bigg|_{\lambda_{1,2}=\lambda_{1,2}^*(\eta,w_2,t)}=0,\\
\dfrac{\partial f(\lambda_1,\lambda_2,\eta,w_2,t)}{\partial \lambda_2}\bigg|_{\lambda_{1,2}=\lambda_{1,2}^*(\eta,w_2,t)}=0.
\end{align}
For a given value of $\eta$, we compute the saddle point
($\lambda_1^*(\eta,w_2,t),\lambda_2^*(\eta,w_2,t))\in
\mathbb{R}_1-\mathbb{R}_2$ where $g(\lambda_1,\lambda_2)$ is analytic,
at fixed $\theta,\ \alpha$, and $t$ as a function of $w_2$. Further, we
compute the integral given in \eref{pw_1w_2-num},
numerically. Finally, the numerical expression for $P_t(-\eta
W_2,W_2)$ is utilized to compute the distribution of efficiency for a
given efficiency $\eta$.

\subsection{Numerical simulation: Possibility I}
\label{result-2}
We compare the analytical results given by \eref{eqn2} with the numerical simulation for parameters $\theta=2.0$ and $\alpha=0.5$ at time $t/t_\gamma=10$. 
\begin{figure}
\includegraphics[width=1.0\linewidth]{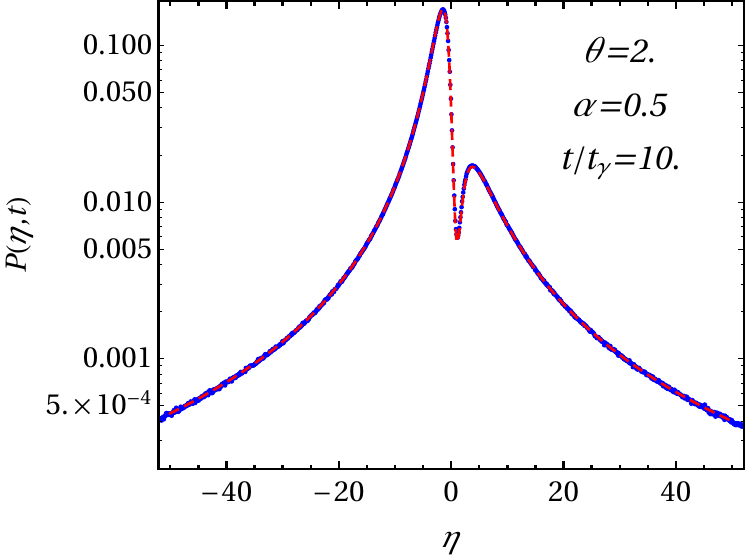}
\caption{Distribution of stochastic efficiency $P(\eta,t)$ is plotted against the stochastic efficiency $\eta$ for $\theta=2$ and $\alpha=0.5$ at time $t/t_\gamma=10$.  Red dashed lines plotted for analytical expression given in \eref{eqn2} while the blue dots are for the numerical simulation. }
\label{dist-phase-21}
\end{figure}
Note that the point $(\theta,\alpha)$ we have chosen lies in the
region (see \frefss{phase-dig-1} and \frefs{phase-dig-3}) where
$g(\lambda_1,\lambda_2)$ has singularities. \ffref{dist-phase-21} shows very good agreement between numerical simulation and
theoretical prediction.

\subsection{Numerical simulation: Possibility II}
\label{result-3}
We compare the numerical simulation for the distribution of stochastic
efficiency with the result obtained by the numerical saddle-point
integration explained in \sref{num-saddle}, for $\theta=3.0$ and
$\alpha=0.5$ at time $t/t_\gamma=50$.
\begin{figure}
\includegraphics[width=1.0\linewidth]{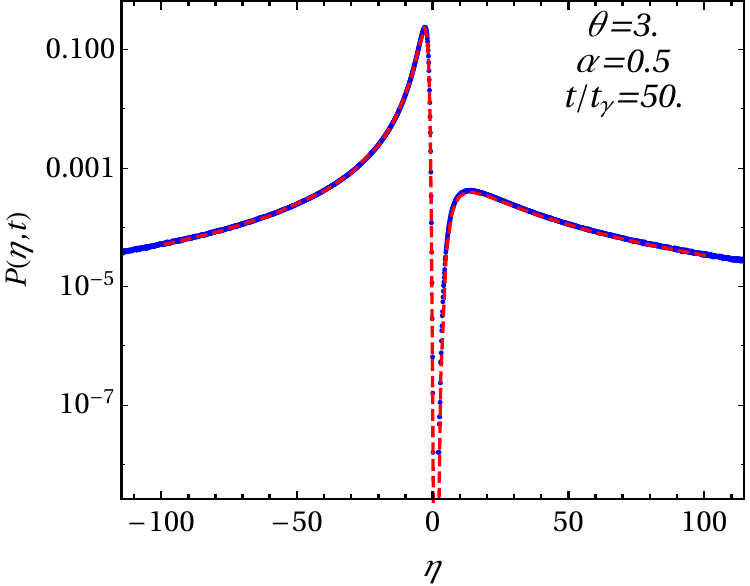}
\caption{Distribution of stochastic efficiency $P(\eta,t)$ is plotted against the stochastic efficiency $\eta$ for $\theta=3$ and $\alpha=0.5$ at time $t/t_\gamma=50$. The red dashed line is plotted  for the distribution $P(\eta,t)$ evaluated numerically as explained in \sref{num-saddle} while the blue dots correspond to the numerical simulation.}
\label{dist-phase-22}
\end{figure}
\ffref{dist-phase-22} shows that there is nice agreement between numerical simulation and theoretical prediction.

\section{Summary}
\label{summ}
We have considered a microscopic
engine in which a Brownian particle is coupled to a heat bath at a
constant temperature, and two external time-dependent forces, called
load force and drive force, are applied to the particle.  Both forces are assumed to be uncorrelated and stochastic Gaussian noises.
The function of the drive force is to drive the Brownian particle against
the load force. Work done by the load force and the drive force, $W_1$
and $W_2$, respectively, is stochastic quantities. Hence, the
efficiency of the engine, which is the ratio of output work to the
input work, $\eta=-W_1/W_2$, is also a stochastic quantity. In this
paper, we have computed the distribution of stochastic efficiency
$P(\eta,t)$ for large $t$.

To compute $P(\eta,t)$, we have first computed the characteristic
function $\langle e^{-\lambda_1 W_1 - \lambda_2 W_2}\rangle \sim
g(\lambda_1,\lambda_2)\ e^{(t/t_\gamma)
  \mu(\lambda_1,\lambda_2)}$. The asymptotic form of the joint
distribution $P_t(W_1, W_2)$ for large $t$ is usually obtained by
inverting the characteristic function using a saddle-point
approximation. We have found that $g(\lambda_1,\lambda_2)$ can have
singularities within the domain where the saddle point lies, and in
that case we have computed the asymptotic distribution $P_t(W_1, W_2)$
by taking the singularities into account. Whether
$g(\lambda_1,\lambda_2)$ has singularities or not depends on the choice
of the parameters $\theta$ and $\alpha$ (see \fref{phase-dig-1}),
which describe the strengths of the external forces relative to each
other as well as to the strength of the thermal noise.

Using $P_t(W_1,W_2)$, we have finally computed $P(\eta,t)$, which have
the large deviation form $P(\eta,t)\sim \exp[(t/t_\gamma)\, J(\eta)]$.
The large deviation function $J(\eta)$ shows two extrema: a minimum
$\eta^*$ corresponds to \emph{an analog of Carnot efficiency} while
the maximum $\bar{\eta}$ is at the most probable efficiency.

As a final remark, since the random external forcing can be realized
in an experimental setup~\cite{Gomez-main}, it would be interesting to
compare the theoretical results obtained here with experiments.


\appendix
\label{sing-contour}
\section{Singularity contour}

The singularity contour given in \eref{red-contour}, can be written in parametric representation as
\begin{eqnarray}
\lambda_1(\Phi)&=&\dfrac{1}{1+\theta+\theta \alpha^2}\bigg(1+\sqrt{1+\alpha^2 }\cos \Phi\bigg) \nonumber\\
\lambda_2(\Phi)&=&\dfrac{1}{1+\theta+\theta \alpha^2}\bigg(1+\dfrac{\sqrt{1+\alpha^2 }}{\alpha}\sin \Phi\bigg)
\label{contour-sing}
\end{eqnarray}
where $\Phi\in(\Phi_-,\Phi_+)$.

When singularity contour given by \eref{red-contour}, intersects the boundary of domain $\mathbb{R}_1$, we get,
\begin{align}
f^-(\lambda_1(\phi_\pm),\lambda_2(\phi_\pm))&=0,\label{eqn1}\\
\cos\phi_\pm+A \sin\phi_\pm-B&=0,
\label{sing-boundary}
\end{align}
where \begin{align}
A&=\alpha\sqrt{1+\theta+\theta \alpha^2},\\
B&=\sqrt{(1+\alpha^2\theta)/[\theta (1+\theta+\theta \alpha^2)]}(1-\theta-\theta \alpha^2).
\end{align}
In \eref{sing-boundary}, we have used $\nu\big(\lambda_1(\phi_\pm),\lambda_2(\phi_\pm)\big)=0$.

\begin{figure}[!h]
\includegraphics[width=1.0\linewidth]{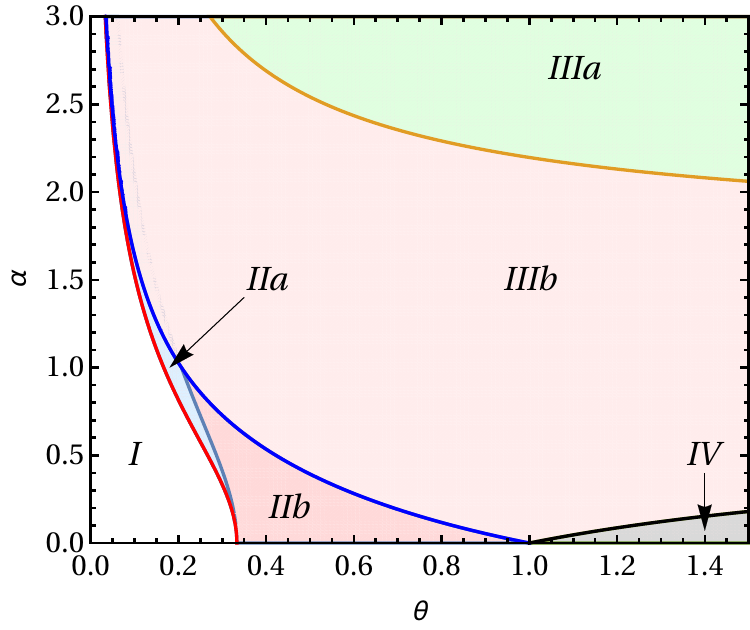}
\caption{Phase diagram indicates the sign of $C_{1,2}^\pm$. Region I: $g(\lambda_1,\lambda_2)$ is analytic. In subregions IIa: $C_{1,2}^\pm>0$. In subregion  IIb: $C_{1,2}^+>0$, $C_1^-<0$ and $C_2^->0$. In subregion IIIa: $C_{1,2}^+<0$, $C_1^-<0$ and $C_2^->0$. In subregion IIIb: $C_1^+>0$, $C_2^+<0$ , $C_1^-<0$ and $C_2^->0$. In region IV: $C_1^+>0$, $C_2^+<0$  and $C_{1,2}^-<0$. }
\label{modified-phase-dig-1}
\end{figure}
Consider $\sin \phi_\pm=x$; therefore \eref{sing-boundary} becomes
\begin{equation}
\pm \sqrt{1-x^2}=-A x+B.
\label{x-pm}
\end{equation}
The solution of \eref{x-pm} is given by
\begin{equation}
x_\pm=\dfrac{AB\pm\sqrt{1+A^2-B^2}}{A^2+1}.
\end{equation}

Since $x_\pm$ is a real number, therefore, $1+A^2-B^2=-1+3(1+\alpha^2)\theta\geq0$. Thus,
\begin{equation}
-1+3(1+\alpha^2)\theta\geq0
\end{equation}
gives us the restriction on $\alpha$ and $\theta$ for which the singularity contours appears in the scenario as shown in \fref{contours}(b). Using above inequality, we have plotted the phase diagram shown in \fref{phase-dig-1}.
$x_\pm$ is the solution of equation $+\sqrt{1-x^2}=-A x+B$ when 
\begin{align}
&(1) \quad\quad B\geq +A\sqrt{1+A^2-B^2} \quad\quad\quad \text{for}\quad\quad\quad x_+,\\
&(2) \quad\quad  B\geq -A\sqrt{1+A^2-B^2} \quad\quad\quad\text{for}\quad\quad\quad x_-.
\end{align}
Similarly, $x_\pm$ is the solution of equation $-\sqrt{1-x^2}=-A x+B$ when 
\begin{align}
(3) \quad\quad B&\leq +A\sqrt{1+A^2-B^2} \quad\quad\quad \text{for}\quad\quad\quad x_+,\\
(4) \quad\quad B&\leq-A\sqrt{1+A^2-B^2} \quad\quad \quad\text{for}\quad\quad\quad x_-.
\end{align}
Therefore, using $x_\pm$ and conditions $(1)-(4)$, one can find the end points of the contour $\lambda_{1,2}(\phi_\pm)$.

\subsection{Range of $\Phi$}
Comparing $\lambda_{1,2}(\Phi_\pm)=\lambda_{1,2}(\phi_\pm)$, we get
 \begin{align}
\sin \Phi_\pm=C_1^\pm \quad\quad \text{and} \quad\quad \cos \Phi_\pm=C_2^\pm,
\label{end-eqn}
\end{align} 
where 
\begin{align}
C_1^\pm&=\dfrac{\alpha}{\sqrt{1+\alpha^2}}[(1+\theta+\theta\alpha^2)\lambda_2(\phi_\pm)-1],\\
\quad C_2^\pm&=\dfrac{1}{\sqrt{1+\alpha^2}}[(1+\theta+\theta\alpha^2)\lambda_1(\phi_\pm)-1],
\end{align}
with $(C_1^\pm)^2+(C_2^\pm)^2=1$.
Using \eref{end-eqn}, one can find the restriction on $\Phi$, which is given as
\begin{equation}
\Phi_\pm=-i \ln[C_2^\pm+i  C_1^\pm].
\end{equation}
The sign of $C_{1,2}^{\pm}$ can be anything. Based on the sign, it is decided in which quadrant $\Phi_\pm$ are. Depending upon the sign, we modified the phase diagram \fref{phase-dig-1} as shown in \fref{modified-phase-dig-1}.

Given $\Phi_\pm$, one can use \eref{contour-sing} to plot the singularity contour in $(\lambda_1,\lambda_2)$ plane. It is important to note that the sense of direction is always taken as $\Phi_-$ to $\Phi_+$ (anti-clockwise).
Therefore, the end points of the singularity contour are given by $\lambda_{1,2}^{\pm}=\lambda_{1,2}(\Phi_\pm)$.

\bibliography{ref}
\end{document}